\documentclass[twocolumn,aps,prb,superscriptaddress]{revtex4-2}
	\usepackage{chngcntr}
	\usepackage{graphicx}
	\usepackage{color}
	\usepackage{amsmath}
	\usepackage{enumitem}
	\usepackage{amssymb}
	\usepackage{cancel}
	\usepackage{ulem}
	\usepackage{multirow}
    \usepackage[colorlinks=true,linkcolor=blue,citecolor=blue]{hyperref}

	\newcommand{\be}{\begin{equation}}
		\newcommand{\ee}{\end{equation}}
	
	\newcommand{\bea}{\begin{eqnarray}}
		\newcommand{\eea}{\end{eqnarray}}

	\newcommand{\p}{\partial}
	
	\newcommand{\la}{\left\langle}
	\newcommand{\ra}{\right\rangle}
	\newcommand{\lb}{\left[}
	\newcommand{\rb}{\right]}
	\newcommand{\lp}{\left(}
	\newcommand{\rp}{\right)}

	\renewcommand{\vec}[1]{{\boldsymbol #1}}

\DeclareMathOperator\arctanh{arctanh}
\usepackage[dvipsnames]{xcolor}

	\usepackage{bbold}

\begin{document}

\title{Chirality-induced pseudo-magnetic fields, flat bands and enhancement of superconductivity}
\author{Zhiyu Dong}
\affiliation{Department of Physics, Massachusetts Institute of Technology, Cambridge, MA 02139}
\affiliation{Department of Physics and Institute for Quantum Information and Matter, California Institute of Technology, Pasadena, California 91125}
\author{Leonid Levitov}
\affiliation{Department of Physics, Massachusetts Institute of Technology, Cambridge, MA 02139}
\author{Patrick A. Lee}
\affiliation{Department of Physics, Massachusetts Institute of Technology, Cambridge, MA 02139}

\begin{abstract}
Systems in which exchange interactions couple carrier spins to a spin texture with a net chirality exhibit a spin-dependent Aharonov-Bohm effect, where the geometric gauge field and 
pseudo-magnetic field have opposite signs for carriers with opposite spins. As a result, 
Cooper pairs see a net zero vector potential and superconducting pairing is not hindered by pair-breaking effects. This allows superconductivity to occur even when the geometric field induces quantized Landau levels. We identify the dominant pairing order as an s-wave pair density wave of an FFLO type. 
Flat Landau levels can significantly enhance superconducting $T_c$, favoring superconductivity over competing orders. 
This exotic paired state 
features tell-tale signatures such as flat bands of Bogoliubov-deGennes quasiparticles, 
manifest through Landau level-like resonances 
in the quasiparticle density of states. 
\end{abstract}
\maketitle

\begin{figure}[t]
    \centering
    \includegraphics[width=0.87\linewidth]{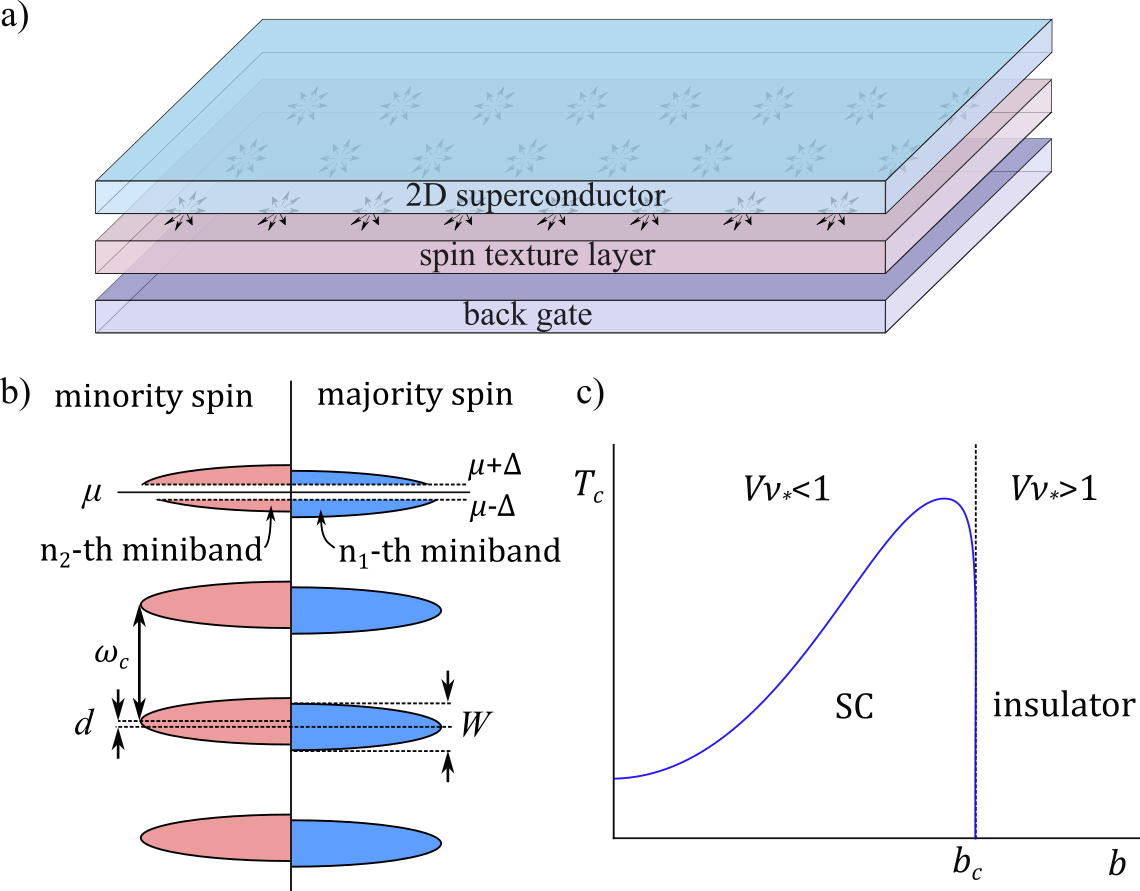}
    \caption{a) Superconducting 2D metal  interfacially coupled to a
    chiral spin texture in a magnetic insulator.
    b) Majority-spin and minority-spin bands in the metal in which  carrier spins are collinear and anticollinear to spin polarization of the texture, respectively. 
    Each band is split into Landau levels (LLs) due to the chiral magnetic field $b$ of the texture. The field $b$ is of opposite signs for spin-up and spin-down carriers. This eliminates the pair breaking effect, whereas LLs enhance the DOS at the Fermi level, assisting SC. 
c) Phase diagram for superconducting $T_c$ vs. $b$.
    }
    \vspace{-6mm}
    \label{fig1:combo}
\end{figure}

Magnetic ordering is commonly viewed as antagonistic to superconductivity, as broken time-reversal symmetry (TRS) results in pair breaking effects that suppress superconductivity (SC) \cite{tinkham2004introduction, de2018superconductivity}.
Here we consider 
magnetic metals in which carrier spins are coupled by exchange interactions to a spin texture with net chirality, as in Refs. \cite{
onoda2002topological,ye1999berry, chen2014anomalous}. 
We predict that SC in such systems can exhibit a fundamentally different behavior.
Chiral textures can generate spin-dependent magnetic fields with opposite signs for carriers of opposite spin.
 Rather than causing pair breaking, such chiral fields facilitate pairing and enhance SC. 
Here we present a theoretical framework and identify 
key signatures of this pairing mechanism.
Chiral spin-textured phases can be classified, perhaps somewhat loosely, into `intrinsic' and `extrinsic'. Intrinsic textures can be part of an itinerant magnetic order in the electron band of interest, as in 
\cite{dong2024chiral, PhysRevLett.101.156402}, or 
in a different (narrow) carrier band \cite{nagaosa2012gauge}.
Extrinsic textures, which will be our focus here, can be induced by 
an adjacent magnetic layer to which spins in a two-dimensional 
metal are coupled by an interfacial exchange coupling  \cite{bruno2004topological,matsuno2016interface,paul2023giant}. 
In such systems 
carriers with opposite spins are split in energy, forming magnetic Bloch bands (see Fig. \ref{fig1:combo}). 
Typically, exchange splitting is assumed much larger than the bandwidth \cite{nagaosa2012gauge}. Here, instead, we focus on an intermediate regime where exchange splitting creates partial spin imbalance, leaving both spin-polarized bands populated.




Here we consider extrinsic textures in coordinate space, deferring the analysis of other textures till later. Non-coplanar spin textures feature nonzero spin chirality $\chi(\vec r)=\la \vec s(\vec r)\cdot\lb \p_x \vec s(\vec r)\times\p_y \vec s(\vec r)\rb \ra $, with $\vec s(\vec r)$ representing the spin density in the continuum limit. 
In the adiabatic regime, exchange interaction locks electron spins to  $\vec s(\vec r)$, whereas the twisting spin quantization axis generates a geometric magnetic field perpendicular to the plane\cite{nagaosa2012gauge}: 
\be\label{eq:b=nabla times a}
b
=\hat{\vec {z}} \cdot \lp \nabla\times \vec a \rp 
= \frac{\phi_0}{4\pi}\chi(\vec r)
,
\ee
Here, $\vec a(\vec r)$ is a geometric gauge field encoding the Berry phase for adiabatic carrier transport with spin aligned with the texture's polarization, with $b$ referred to as the chiral magnetic field. We focus on fields with nonzero spatial averages, requiring topologically nontrivial spin textures like skyrmion crystals. For simplicity, we start with uniform $b$, producing sharp LLs, and address magnetic bands with non-uniform $b$ later.

The adiabaticity condition required for our analysis to be valid 
is that the kinetic energy cost to form a wave-packet of size $\ell_0$ is less than the energy gained from the exchange energy, where $\ell_0$ is the periodicity of spin texture. (We define the length scale $\ell_0$ so that the unit cell area of a skyrmion crystal is $\ell_0^2$, in which case the chiral magnetic field equals $b = \phi_0 /l_0^2$.) For an electron gas at $k_F \ell_0 
\gg 1$, the adiabaticity condition is \cite{metalidis2006topological}
\be\label{eq:condition1}
J\gg \hbar v/\ell_0
,
\ee
where $v=k_F/m$ is the Fermi velocity and $J$ is the carrier-to-texture spin exchange coupling
(for a detailed analysis, see \cite{matsui2021skyrmion}).
Exchange coupling $J$ of carrier spins and the texture spins produces 
the $+$ and $-$ bands (or majority and minority bands): 
$
\epsilon_\pm(p)=\epsilon(p)\mp\frac12 J
$. 
The condition that both bands are partially filled requires $v k_F > J$, which is consistent with Eq.\eqref{eq:condition1}.


A crucial aspect of adiabatic transport that will be key for our analysis of SC 
stems from the fact that carriers with spins aligned parallel or antiparallel to the local texture experience Berry phases of opposite signs. 
Indeed, since the chirality is odd under time-reversal symmetry,
$\vec s(\vec r)\to -\vec s(\vec r)$, 
the $+$ and $-$ carriers couple to
the geometric vector potential with opposite gauge charges. 
Specifically, the Aharonov-Bohm phase, $\exp[i\int dr q_{\pm}\vec a(r)]$, has opposite signs for spin-up and spin-down carriers, with $q_+=-q_-=e$. The gauge charge $e$ is chosen so that the flux of the chiral field matches the flux quantum $h/e$.


We consider SC 
driven by a short-range attraction that, without $b$ fields, would produce a conventional s-wave pairing of opposite spins. With $b$ fields, this becomes a pairing problem for the $+$ and $-$ carrier bands. 
In this case, the $4\times 4$ Bogoliubov-deGennes equations for spin-up and spin-down carriers coupled to the gauge field reduce to two independent $2\times 2$ problems: 
\be\label{eq:BdG}
H=\lp\begin{array}{cc} \epsilon_+(\vec p-q_+ \vec a) -\mu & \Delta (\vec r)
\\
\Delta^* (\vec r) & -\epsilon_-(-\vec p-q_- \vec a) +\mu
\end{array}\rp
.
\ee
%
Importantly, since the gauge charges in two spin flavors are of opposite signs,
the Cooper pairs formed by majority-spin and minority-spin fermions will see a vanishing total geometric magnetic field. As a result, the $b$ field will have no pair breaking effect on SC even when this field is strong or even quantizing. Further, the high DOS in $b$-field LLs or relatively flat bands may lead to $T_c$ enhancement. 
In comparison, for 
conventional magnetic field, where $q_+=q_- = e$, it has been proposed that SC may have a re-entrant state for fields  $B$ larger than $H_{c2}$ when only a few LLs are occupied and Zeeman splitting is negligible\cite{tevsanovic1989quantum, rajagopal1991linearized}. 
In this case, the Cooper pairs have a total gauge charge of $2e$, resulting in vortices  appearing at a high density on the scale of the magnetic length. 
To the contrary, the SC order induced by a quantizing chiral $b$-field 
when LLs align with the Fermi level is vortex-free. 

A striking signature of SC in a chiral magnetic field is the fact that the LL structure in the quasiparticle spectrum survives  the presence of the pairing gap. This behavior is distinct from that of a conventional type II SC, where the presence of a pairing gap suppresses LLs\cite{norman1996absence}. 
The difference arises because Bogoliubov quasi-particles, described by $\gamma_{\uparrow}=u c_{\uparrow}+v c^\dagger_{\downarrow}$, are not minimally coupled to the physical gauge field $\vec A$. In contrast,
for chiral magnetic field, $c_{\uparrow}$ and $c^\dagger_{\downarrow}$ carry the same gauge charge, rendering the pairing order parameter $\Delta$ charge neutral. Hence $\gamma$  carries a well defined gauge charge and the quasi-particle spectrum remains discrete in the SC phase. 

The LL quasiparticle spectrum can be probed in several ways. One way is to measure quantum oscillations of magnetization. Unlike the usual quantum oscillations, here, we expect the peak of magnetization quantum oscillation to be widened due to the presence of pairing gap at Fermi level. However, the periodicity of quantum oscillations  
vs. carrier density 
will match the normal LLs. The discrete LLs in quasiparticle spectrum can also be probed by optical or scanning tunneling spectroscopy. 




We note that the enhancement of $T_c$ via the enhanced density of states (DOS) of flat bands, which has received much attention in the past, is a delicate question. In general, the enhanced DOS also enhances other instabilities such as spin or charge density waves or ferromagnetism. However, for LLs in a chiral magnetic field, 
the order parameter for spin density waves $\langle c_+^{\dagger}(\vec k+\vec p)  c_-(\vec k) \rangle$ carries gauge charge $2e$ and is suppressed in the same way as SC would be suppressed by physical magnetic field. Other types of density waves are analyzed in Appendix \ref{app:competing instabilities}.

Another important point is that the leading pairing instability in an exchange-split 
Fermi sea 
is expected to occur in a pair-density wave (PDW) channel.
PDW phases are of course well known for superconducting metals where spin-up and spin-down Fermi seas are split by 
spin effects such as spin-orbital coupling or Zeeman coupling to an external $B$ field. At intermediate field strengths, these interactions result in a 
PDW state of a FFLO type  \cite{fulde1964superconductivity, larkin1964nonuniform}, followed by a complete suppression of SC at higher fields. 
Superconducting PDW phases feature intriguing non-reciprocal supercurrent response
\cite{watanabe2022nonreciprocal}.
For the scenario considered here, described by the Fermi sea splitting and the associated PDW, the main effect is the formation of geometric LLs in each of the spin-split Fermi seas. This creates flat 
quasiparticle bands, with SC enhanced when these bands align with the Fermi level.

Turning to the analysis, we consider the problem $H=H_0+H_{\rm int}$ with an attractive pairing interaction 
\be
H_{\rm int}=-\!\! 
 \sum_{\vec r,\vec r', s,s'} \!\!\frac{g(\vec r - \vec r')}2 
\psi^\dagger_s (\vec r)\psi^\dagger_{s'} (\vec r')\psi_{s'} (\vec r')\psi_{s} (\vec r)
.
\ee
The term $H_0$ describes spin-split bands of carriers exchange-coupled to  
spin texture:
\be
H_0 = \sum_{\alpha=\pm} \psi^\dagger_{\alpha}(\vec r)\lb \epsilon_\alpha\lp -i\nabla-q_\alpha \vec a(r)\rp-\mu \rb \psi_\alpha(\vec r)
\ee
where the geometric gauge field $\vec a$ is associated with the chiral magnetic field $b$, Eq.\eqref{eq:b=nabla times a}. 
For a parabolic band, $H_0$ is diagonalized in terms of LLs in the usual way, giving 
\be\label{eq:external_texture_energy}
\epsilon_{n}^{\pm} = \epsilon_n \mp J/2 = \lp n+1/2 \rp \hbar \omega_c \mp J/2 -\mu
,
\ee
where $\omega_c=|q_+|b/m$ is the cyclotron frequency due to the chiral gauge field $b$. 
Here $\epsilon^\pm_{n}$ represents the energy of the $n$-th LL in $\pm$ flavors measured from the Fermi level. 
The Bogoliubov quasi-particle energies are given by
\be
E_{n_1 n_2}= 
\tilde{\epsilon}_{n_1 n_2} \pm \sqrt{\Delta^2+\bar{\epsilon}_{n_1 n_2}^2 } 
\ee
where $n_1$ and $n_2$ are the indices of the two aligned levels in $+$ and $-$ flavors between which pairing occurs. Here, we defined $\bar{\epsilon}_{n_1 n_2} = \frac1{2}\lp \epsilon_{n_1}^+ + \epsilon_{n_2}^- \rp$ and $\tilde{\epsilon}_{n_1 n_2} = \frac1{2}\lp \epsilon_{n_1}^+ - \epsilon_{n_2}^- \rp$. Interestingly, the quasiparticle spectrum, which is equally spaced in the normal state, is no longer so in the presence of the pairing gap [see Fig.\ref{fig1:combo} (b)].

Next, we proceed to analyze the pairing instability. We focus on the inter-flavor pairing channel $\Delta(\vec r_1,\vec r_2)\sim \langle \psi_+(r_1) \psi_- (r_2)\rangle $ in which
conventional mechanisms, such as a phonon-mediated attraction, generate an even-parity pairing. In contrast, pairing in intra-flavor channel, which could yield an odd-parity pairing gap function,  is unlikely under conventional pairing mechanisms 
\cite{footnote1}.




The inter-flavor pairing vertex is governed by the following self-consistency equation
\be \label{eq:Delta1}
\Delta(\vec r_1,\vec r_2) = \int d\vec r_1'd\vec r_2' 
K(\vec r_1,\vec r_2;\vec r_1',\vec r_2') 
\Delta(\vec r_1',\vec r_2'). 
\ee
Here the kernel is 
$K(\vec r_1,\vec r_2;\vec r_1',\vec r_2') 
= -g(\vec r_1 -\vec r_2) \sum_\omega G_+(\omega; \vec r_1,\vec r_1') 
G_-(\omega; \vec r_2,\vec r_2')$, 
where $G_\pm(\omega;\vec r,\vec r')$ are the 
Green's functions for fermions of $\pm$ flavor 
\begin{align}
G_\pm(\omega;\vec r,\vec r') = -\langle \mathcal{T} \psi_{\pm}(\omega;\vec r)\psi_{\pm}^\dagger(\omega;\vec r')\rangle
=  \sum_{n} \frac{ A_{n,\pm}(r,r')
}{i\omega - \epsilon_{n}^{\pm}}, 
\nonumber 
\end{align}
with $A_{n\pm}(r,r') = \sum_m \psi_{nm\pm}(\vec r)\psi^*_{nm\pm}(\vec r')$, $\psi_{nm\pm}(\vec r)$ are the wavefunctions of $m$-th degenerate state in the $n$-th LL in the $\pm$ flavor.

Plugging in the LL wavefunctions, 
we obtain 
an expression for the kernel:
\begin{align}
& K(\vec r_1,\vec r_2;\vec r_1',\vec r_2') =\frac1{2}\sum_{n_1,n_2}\frac{\tanh \frac{\epsilon_{n_1}^{+}}{2T} + \tanh \frac{ \epsilon_{n_2}^{-}}{2T} }{\epsilon_{n_1}^{+} +\epsilon_{n_2}^{-} } \lp\frac{1}{2\pi \ell_b^2}\rp^2\nonumber \\
& \times M_{n_1,n_2}e^{-(r_{11'}^2 +r_{22'}^2)/4\ell_b^2 } e^{-i \lp \vec r_1 \times \vec r_1' - \vec r_2 \times \vec r_2' \rp \cdot \hat{\vec z}/2\ell_b^2 }
\end{align}
where 
$M_{n_1,n_2}=L_{n_1}\lp \frac{r_{11'}^2}{2\ell_b^2}\rp L_{n_2}\lp\frac{r_{22'}^2}{2\ell_b^2}\rp 
$, $n_1$ and $n_2$ label the LL index of the $\pm$ bands. Here 
we defined a magnetic length $\ell_b = \sqrt{\frac{\phi_0}{2\pi b}} = \frac{\ell_0}{\sqrt{2\pi}} $
and displacements $r_{11'} = r_{1}-r_{1'}$ and $r_{22'} = r_{2}-r_{2'}$.  The minus sign in the phase factor reflects the opposite signs of gauge charges in $\pm$ favors. 

It is instructive to compare this expression 
to that for the kernel in the problem of pairing in a physical magnetic field, studied in Refs.\cite{tevsanovic1989quantum,rajagopal1991linearized}. In that setting, the kernel takes a similar form, with a crucial difference that the phase factor in the kernel takes the form of $r_1\times r_1'+ r_2\times r_2'$  \cite{tevsanovic1989quantum,rajagopal1991linearized}. The extra minus sign in the phase factor in our problem is crucial because, as we will see below, it implies the center-of-mass motion of a short-range Cooper pair does not sense any gauge field. 
In comparison, in the physical magnetic field problem, Cooper pair wavefunction in the $xy$ plane is localized 
\cite{tevsanovic1989quantum,rajagopal1991linearized} since pairing occurs on top of localized Landau orbitals. In that setting, vortices are generated on the scale of the magnetic length in the $xy$ plane, whereas in our setting there are no vortices.

To see this explicitly, we define the center-of-mass coordinate 
and a relative coordinate 
$
\vec R = (\vec r_1+\vec r_2)/2$, $\vec r = \vec r_1 - \vec r_2 
$.
In this notation, the gap equation yields
\begin{align}
 \label{eq:gap_eq}
 &\Delta(\vec R;\vec r)  = -\frac{ g(r) }{2(2\pi \ell_b^2)^2}
 \sum_{n_1,n_2} \frac{\tanh \frac{\epsilon_{n_1}^{+}}{2T} + \tanh \frac{ \epsilon_{n_2}^{-}}{2T} }{\epsilon_{n_1}^{+} +\epsilon_{n_2}^{-} } \\
& \times \int M_{n_1,n_2} e^{-\frac{\delta R^2}{2\ell_b^2} - \frac{\delta r^2}{8\ell_b^2} -i \frac{\lp \delta \vec r \times \vec R - \delta \vec R \times \vec r \rp \cdot \hat{\vec z}}{2\ell_b^2} }  \Delta(\vec R+\delta \vec R; \vec r +\delta \vec r) \nonumber
\end{align}
where integration is carried out over $\delta\vec r$ and $ \delta\vec R$. 
To proceed analytically, we consider a short-range attractive interaction
$
g(\vec r) = g\delta(\vec r)$.
Under this interaction, the form of gap function will also be local
$
\Delta(\vec R;\vec r) = \Phi(\vec R) \delta(\vec r)
$.
Due to the absence of time-reversal or inversion symmetries, we generally expect the leading pairing instability occurs in a finite momentum channel. Therefore, we consider a pair-density-wave channel:
\be\label{eq:PDW}
\Phi(\vec R) = \Phi_0 e^{i\vec P\cdot \vec R}
\ee
Mathematically our problem is equivalent to that in the Gor'kov and Dzyaloshinskii's analysis of the particle-hole pair wavefunction of an exciton in a magnetic field \cite{gor1968contribution}, showing that the dependence on the center of mass coordinate $\vec R$  can 
generally be factored out as $e^{i\vec P\cdot \vec R}$ as in Eq.\ref{eq:PDW}. This follows from the conservation of total momentum due to the opposite gauge charges. Therefore, this factorization is a general result that does not  rely on the delta function form of the interaction. 





Plugging Eq.\eqref{eq:PDW} into Eq.\eqref{eq:Delta1}, yields the gap equation 
\begin{align}
& 1 = \frac{g}{2\pi \ell_b^2} \sum_{n_1,n_2} \frac{\tanh \frac{\epsilon_{n_1}^{+}}{2T} + \tanh \frac{ \epsilon_{n_2}^{-}}{2T} }{\epsilon_{n_1}^{+} +\epsilon_{n_2}^{-} } I_{n_1n_2}(p),
\label{eq:linearized_gap_eq_PDW} \\
& I_{n_1n_2}(p) = \int \frac{d^2 r}{2\pi}  L_{n_1}\lp r^2\rp L_{n_2}\lp r^2\rp e^{-r^2 }  e^{i p r \cos\theta} 
\label{eq:def_I}
\end{align}
where we have defined dimensionless momentum $p = \sqrt{2} \ell_b P$ and dimensionless distance $ r = R/(\sqrt{2}\ell_b)$. 
Numerical analysis of the structure factor $I(p)$ (see Appendix \ref{app:Tc in flat Landau level}) shows that it is maximized at $p \sim \sqrt{n_1} - \sqrt{n_2}$ for any given $n_1$ and $n_2$, implying that the leading PDW order occurs at a momentum of $P\sim (\sqrt{n_1} -\sqrt{n_2})/\ell_b$. This is consistent with the familiar FFLO result in the absence of magnetic field where PDW momentum is given by $P = k_{F1} -k_{F2}$, where $k_{F1}$ and $k_{F2}$ represent the Fermi momenta in $\pm$ flavors.

Eqs. \eqref{eq:linearized_gap_eq_PDW} and \eqref{eq:def_I} can be employed to study 
enhancement of $T_c$ occurring when pairs of LLs in $\pm$ flavors align with  the Fermi level. 
Ideally, the $T_c$ will be maximized when LLs in two spin flavors are aligned (see Fig.\ref{fig1:combo}). This is achievable by tuning the splitting $J$ by pressure or by applying an out-of-plane electric field. Due to the vanishing bandwidth, the pairing problem becomes essentially a zero-dimensional pairing problem. 
The pairing susceptibility scales as $1/T$ (see Eq.\eqref{eq:linearized_gap_eq_PDW}), leading to a maximum critical temperature of $T_c^{(0)}= 0.17g/4\pi \ell_b^2$ occurring when the lowest LL in minority spin and the second lowest LL in majoirty spin are at the Fermi level (i.e. $n_1=1$ and $n_2=0$, see Eq. \ref{eq:Tc_PDW_1}. Using the dimensionless coupling $\lambda=g\nu$, where $\nu=m/2\pi$ is the bare DOS, we find $T_c^{(0)}= 0.17 \lambda \hbar \omega_c/2$. This $T_c$ value is indeed enhanced upon increasing $b$ and can be fairly large.

However, it turns out that aligning the LLs in two spin flavors 
can become problematic in the presence of ee interactions. As in the Stoner instability problem, a large DOS at the Fermi level is unfavorable and  the exchange interaction between electrons  generates spontaneous spin-splitting field that detunes the LLs in two spin flavors. We define the detuning energy $d$ as the energy difference between LLs of the $\pm$ bands closest to the Fermi levels. As detailed in Appendix \ref{app:detuning}, after accounting for exchange energy, we find that $d$ cannot be tuned to zero. 
In fact, at zero temperature, the minimal detuning one can achieved is $d_{\min} \sim V/(2\pi \ell_b^2)$ (see Appendix \ref{app:Tc in flat Landau level}). This detuning limits the SC enhancement as it gaps out at least one of the two LLs. From Eq.\eqref{eq:linearized_gap_eq_PDW}, we see that this gap cuts off the $1/T$-scaling of SC susceptibility at $1/d$, so that SC is completely suppressed unless $T_c^{(0)}>d$. Comparing $d_{\rm min}$ with the expression of $T_c^{(0)}$, we conclude that achieving SC requires $0.17 g\gtrsim V$, which is in principle possible by adding proximate metallic screening but difficult to achieve in practice.

This analysis indicates that the difficulty of the ideal setting above is  due to the zero width of LLs 
leading to an infinite DOS. If the width $W$ of LLs is finite, the Fermi surface would not be gapped out by a moderate detuning.
We are therefore led to consider 
a modified model where LLs are broadened into minibands of width $W$ due to spatial inhomogenuity or disorder, as illustrated in Fig.\ref{fig1:combo}.  
Detailed analysis presented in Appendix \ref{app:Tc in broadened Landau level} predicts the following maximal $T_c$ in broadened LLs:
\be\label{eq:enhanced Tc}
T_c = W e^{-1/0.34\lambda_*} 
,\quad
\lambda_* = g\nu_*
,
\ee
where $\nu_*$ is the DOS in magnetic bands representing broadened LLs, whereas $W$ is the bandwidth. Similar to the case of narrow LLs, the maximal $T_c$ in broadened LLs is also achieved when $n_1=1$ and $n_2=0$ 
\footnote{This situation occurs in a dilute limit when $k_F \ell_0 \ll1$. In this regime, the adiabaticity condition is given by
$
J\gg \frac{1}{m\ell_0^2 }
$, see Ref.\cite{paul2023giant}.
}.
Since $\nu_*$ grows as $b$ increases, 
$T_c$ can be enhanced to the point where 
$\lambda_*$ approaches order one. At this stage, the system transitions into the strong coupling regime, and 
$T_c$ saturates at $O(W)$. At larger $b$, SC gives way to an insulating state resulting from spontaneous detuning away from LL alignment, 
as illustrated in Fig.\ref{fig1:combo} c). 

The analysis described above was based on a uniform-$b$ model. However, in realistic systems this is usually not the case. 
The skyrmions form a lattice, and the $b$ field generated by spin chirality is spatially modulated. This modulation endows each Landau miniband with a dispersion. If the bandwidth 
becomes comparable to band gaps $\omega_c$ \cite{hamamoto2015quantized}, the observables predicted above can be affected. While the broadening of the DOS can be treated as before for LLs, our assumption that the wavefunctions resemble those of LLs when we compute the matrix elements may no longer be valid.

Notably,  recent work has shown that, some Landau minibands in a metal coupled to skyrmion lattice can be flattened significantly by fine tuning the spin texture. Namely, Refs.\cite{paul2023giant,reddy2024non} report that when the overall spin polarization $\bar{m}$ is tuned to some ``magic" values, the lowest few minibands become as flat as a LL (i.e. smaller than the band gaps by at least an order of magnitude), even though the spin chirality still remains nonuniform. Such a fine tuning of overall polarization $\bar{m}$ is achievable through applying an in-plane magnetic field Zeeman-couples to carriers. Such setting can be naturally included in our model based on LLs, pointing to one promising way to realize the predicted boost in $T_c$.
We note that for a skymion lattice with unit cell area $l_0^2$
, the chiral gauge field is given by $\phi_0/l_0^2
$ and can be large. For example, 
$l_0=5\rm{nm}$
we find $b \approx 160 $T. For carriers with $m=m_e$, this gives a cyclotron energy $\hbar \omega_c=18.5 $meV. The ideal situation will have several LL occupied, which can be achieved with a metal with moderately low density. For example, we can set the Fermi energy $\epsilon_F$ to be of order 100 meV. Using the relation $k_Fl_0=\sqrt{
4\pi\epsilon_F/\hbar \omega_c}$,
we can rewrite the adiabaticity condition Eq.\ref{eq:condition1} and combine with the condition that both $\pm$ bands are partially occupied to give the following condition on $J$:
\be\label{eq:condition3}
\epsilon_F \gg J\gg \epsilon_F\sqrt{\hbar \omega_c/\pi
\epsilon_F}
,
\ee
Using the values $\epsilon_F$ and $\omega_c$ given above, we find a $J$ range of $24$-$100$ meV. These exchange coupling values, determined by the overlap between carrier orbitals and localized spins in the insulator, are realistic \cite{paul2023giant}, matching the scenario pictured in Fig. \ref{fig1:combo}.

Insulating skyrmion materials are relatively rare, with one prominent example being GaV$_4$Se$_8$, where the lattice constant is about 20 nm.\cite{kezsmarki2015neel} The skymion crystal can survive in thin samples where competing helical orders are suppressed.\cite{akazawa2022topological}. Skymion crystals also exist in van der Waals materials that can serve as building blocks for hetero-structures. Examples include Fe${}_2$GaTe$_3$ (metallic) \cite{park2021neel}, its semi-conducting cousin\cite{powalla2023skyrmion} and hybrids \cite{wu2022van}. These have lattice constant $\approx80nm.$ While we have focused on insulating skymion layers in this paper, in principle metallic skymion layer can also work as long as the metallic band is not strongly hybridized with the metal layer, so that the band does not become too wide and the adiabatic condition can still be met. This can be the case when the Fermi surfaces of the layers do not overlap.

We thank Joseph Checkelsky, Jung Hoon Han and Jiaqi Cai for useful discussions.
ZD is supported by the Gordon and Betty Moore Foundation’s EPiQS Initiative, Grant GBMF8682.
PAL 
is supported by DOE (USA) office of Basic Sciences Grant DE-FG02-03ER46076. 
\bibliography{ref}

\begin{thebibliography}{29}
\bibitem [{foo()}]{footnote1}%
  \BibitemOpen
  \href@noop {} {}\bibinfo {note} {We note that an intra-flavor pairing sees
  the chiral magnetic field in the same way as the conventional pairing sees
  the real magnetic field.}\BibitemShut {Stop}%
\end{thebibliography}
\newpage


\appendix

\centerline{\bf Supplementary Material overview}
\medskip 
The following sections provide a detailed derivation for pairing and other instabilities in flat Landau levels and broadened Landau levels.  We calculate $T_c$ for superconducting pairing for flat Landau levels
in Appendix~\ref{app:Tc in flat Landau level}. Next, we derive the conditions for other instabiliities in Appendix~\ref{app:density wave}. This analysis motivates us to focus on spontaneous detuning. In Appendix~\ref{app:detuning}, we work out the detuning in flat Landau levels. This analysi will show that, in flat Landau levels, the pairing is suppressed by detuning. Starting from Appendix~\ref{app:Tc in broadened Landau level}, we switch to study the broadened Landau levels. We work out the pairing Tc and detuning in this setting in Appendix~\ref{app:Tc in broadened Landau level} and Appendix~\ref{app:detuning_broad} respectively. This analysis will show that, detuning does not kill the pairing in this case, although it sets an upper bound of density of states, thus limiting the maximal pairing $T_c$. In Appendix~\ref{app:density wave broad band} we check that other density wave instabilities will not occur, and therefore SC pairing is not obstructed. The conclusion of this analysis is that there is a range of chiral fields $b$ in which $T_c$ is substantially enhanced before superconductivity collapses due to LLs detuning from Fermi level.

\section{PDW $T_c$ in flat Landau levels}\label{app:Tc in flat Landau level}






Below we first allow detunning $d$ to take any value and obtain the maximal $T_c$ of PDW for any given values of $d$. Namely, we will focus on two limits that are easy to calculate: (a) $d=0$, (b) $d\gg T_c$. However, due to the presence of electron-electron interaction, detuning $d$ is not a directly controllable quantity --- $d$ should be determined self-consistently by minimizing energy (including exchange interaction between carriers). Later, we will study what range of $d$ is accessible, and obtain the the maximal $T_c$ accordingly.

\begin{figure}
    \centering
    \includegraphics[width=0.49\columnwidth]{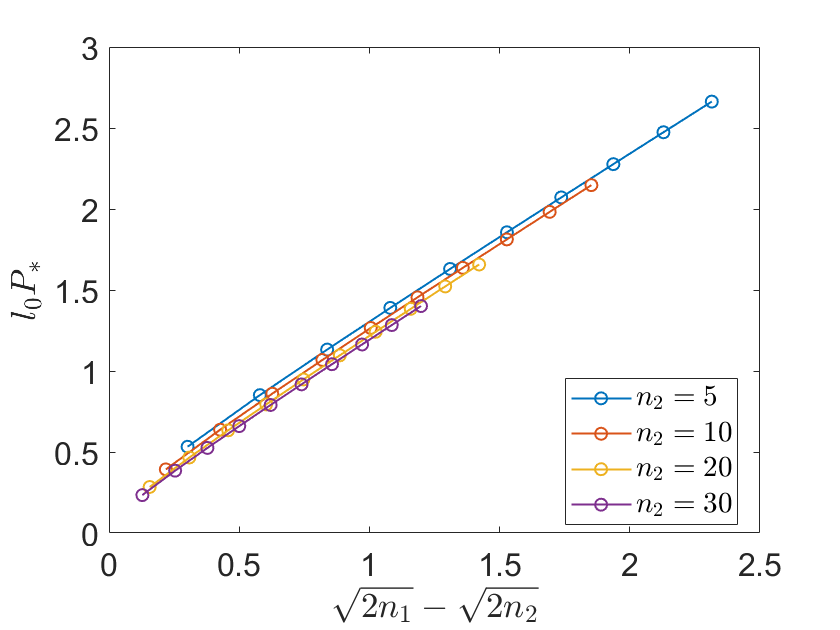}
    \includegraphics[width=0.49\columnwidth]{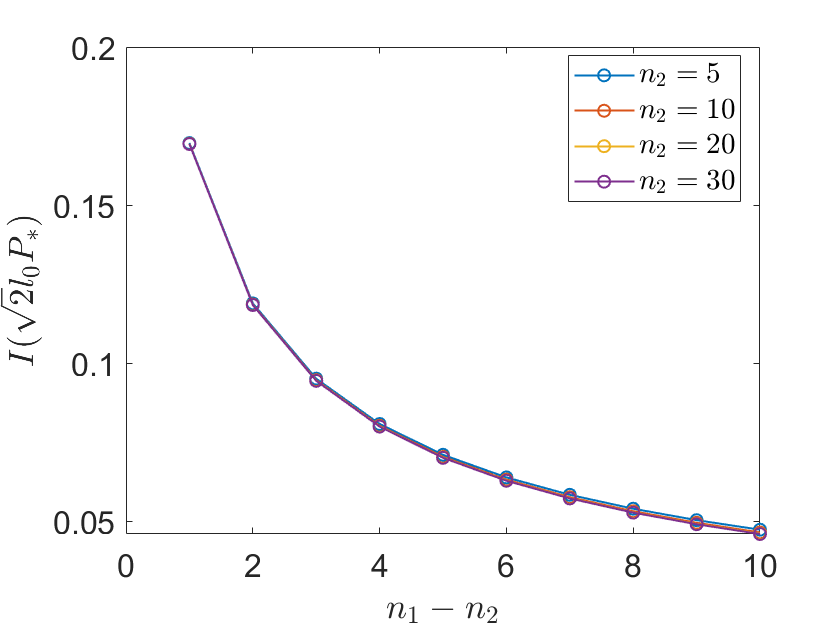}
    \caption{ The maximum of the integral $I_{n_1,n_2}(\sqrt{2} \ell_b P )$ obtained numerically. a) The momentum $P_*$ that maximizes integral $I_{n_1,n_2}(\sqrt{2} \ell_b P )$, b) The maximal value of integral $I_{n_1,n_2}(\sqrt{2} \ell_b P )$ at various values of $n_1$ and $n_2$. }
    \label{fig:numerical integral}
\end{figure}

\underline{\textbf{(a) PDW $T_c$ without detuning ($d=0$):}}
In this case, the LLs in both flavors are simultaneously sufficiently close to Fermi level, i.e. $\epsilon_{n_1}^+= \epsilon_{n_2}^-\ll T$, which is the optimal situation for pairing formation. Plugging this condition into the linearized gap equation Eq.\eqref{eq:linearized_gap_eq_PDW} yields the following expression for $T_c^{(0)}$ of PDW:
\be \label{eq:Tc_PDW}
T_c^{(0)} = \frac{g}{4\pi \ell_b^2} I_{n_1,n_2} (\sqrt{2} \ell_b P),
\ee
Here we have used the definition of dimensionless wavenumber $p = \sqrt{2} \ell_b P$, where $P$ is the PDW wavenumber. Eq.\eqref{eq:Tc_PDW} predicts a leading PDW instability at a momentum $P = P_*$ that maximizes the integral $ I_{n_1,n_2} (\sqrt{2} \ell_b P_* )$.  We analyze the integral $I_{n_1,n_2} (\sqrt{2} \ell_b P) $ numerically, and find that: 

a) For a given choice of $n_1$ and $n_2$, the integral is maximized at roughly $\ell_b P \approx \sqrt{2n_1}-\sqrt{2n_2}$ (see Fig.\ref{fig:numerical integral} a)). This is in agreement with the result at $b=0$, where the FFLO phase occurs at momentum $P = k_{F1} -k_{F2}$. Here $k_{F1}$ and $k_{F2}$ represents the Fermi momentum in two flavors. 

b) The maximal value of this integral is further maximized at $n_1-n_2 = 1$. In this case, the maximal value of $I_{n_1n_2}$ is always around $0.17$ for any value of $n_2$ (see Fig.\ref{fig:numerical integral} 
b) Therefore, the $T_c$ for this leading PDW channel is
\be \label{eq:Tc_PDW_1}
T_c^{(0)} = \frac{0.17g}{4\pi \ell_b^2}.
\ee


\underline{\textbf{(b) PDW $T_c$ under a large detuning $d\gg T_c$:}} In this regime, for a generic filling in the lowest-energy(measured from Fermi level) LLs, one flavor will be pushed away from Fermi level by $\sim d$, whereas the other will stay near Fermi level. without losing generality, we assume the $n_1$-th LL in majority flavor is always near the Fermi level within the window of width $T$, whereas the $n_2$-th LL in minority flavor is detunned from Fermi level by $d$.  the linearized gap equation for optimal PDW channel becomes

\begin{align}
& 1 = 2T_c^{(0)} \frac{\tanh \frac{d}{2T_c} }{ d },\label{eq:linearized_gap_eq_PDW2}
\end{align}
where we have ignored the $\epsilon^+_{n_1}$ terms in nuemrator and denominator of Eq.\eqref{eq:linearized_gap_eq_PDW2} as $\epsilon^+_{n_1} \ll d$, and have used Eq.\eqref{eq:Tc_PDW}.  This yields a Tc of
\be\label{eq:Tc_d>>Tc}
T_c(d) = T_c^{(0)} \frac{d/2T_c^{(0)}}{\arctanh  (d/2T_c^{(0)})}
\ee
At $d=2T_c^{(0)}$, the PDW is completely killed as $T_c$ vanishes. This agrees with our expectation that PDW survives for $d<2T_c^{(0)}$.  As $T_c^{(0)}$ value is enormous, there is sufficient room to tune $d$ without killing PDW. 

We note parenthetically that, strictly speaking, the expression Eq.\eqref{eq:Tc_d>>Tc} becomes invalid for $d<2T_c^{(0)}$ since the $T_c$ predicted by Eq.\eqref{eq:Tc_d>>Tc} equals to $T_c^{(0)}$ which exceed $d$ and violates the assumption of $d\gg T_c$. However, the claim that PDW is completely suppressed at $d=2T_c^{(0)}$ is solid. This is because at $d=2T_c^{(0)}$, Eq.\eqref{eq:Tc_d>>Tc} predicts $T_c=0$ which is consistent with the assumption $d\gg T_c$.

\section{Competing instabilities for flat Landau levels.}
\label{app:competing instabilities}
So far it seems the pairing can always occur at low $T$. However, there might be other orders that competes with SC, such as density waves and spontaneous spin polarization (which is spin density wave at q=0). This is indeed very likely to occur in the LL system because, while the flatness of the LL  enhances the pairing instability, it also enhances instabilities toward other competing orders. 



To start, we observe that there could be two families of density-wave order: (a) the intra-flavor one which is formed by condensing the particle-hole pair in the same flavor, such as $\langle c^\dagger_{k+q,+}c_{k,+}\rangle$ and $\langle c^\dagger_{k+q,-}c_{k,-}\rangle$; and (b) the inter-flavor one, which is formed by condensing the particle in one flavor and hole in the other flavors, such as $\langle c^\dagger_{k+q,-}c_{k,+}\rangle$. Below we argue that the former is suppressed by chiral B field, whereas the latter is not. 

\underline{\textbf{(1) Inter-flavor density wave:}}
The inter-flavor density waves are suppressed in the presence of spin chirality due to the nonvanishing total charge of the condensate. Namely, electrons in $+$ flavor and holes $-$ flavor have a total charge of $2e$. As a result, the inter-flavor density wave, which is the condensate of such particle-hole pairs, behaves like a SC in the presence of a magnetic field. 

\begin{figure}
    \centering
    \includegraphics[width=0.9\linewidth]{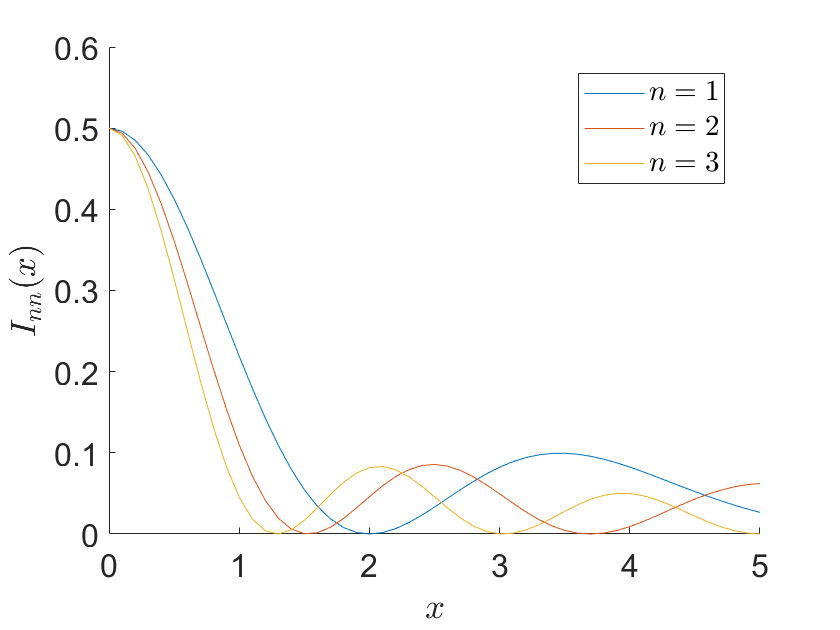}
    \caption{The momentum dependence of structure factor $I_{nn}(x)$ for intra-flavor density wave susceptibility (see Eq.\eqref{eq:Lambda}) }
    \label{fig:I_nn}
\end{figure}

\underline{\textbf{(2) Intra-flavor density wave:}}
In comparison, the intra-flavor density waves can survive in the presence of the chirality since intra-flavor particle-hole pairs have a vanishing total gauge charge. We analyze the intra-flavor density wave instability (detailed in Appendix.\ref{app:density wave}), and find the following criterion
\be
V\tilde{\Pi}(\vec q) = 1,\quad \tilde{\Pi}(\vec q) = \frac{2\Pi_+(\vec q)\Pi_-(\vec q)}{\Pi_+(\vec q) + \Pi_-(\vec q)}.\label{eq:intraflavor_dw_result1}
\ee
where $\Pi_\pm (q)$ represents the total charge susceptibility in LLs, which is given by
\begin{align}\label{eq:Pi+}
&\Pi_+ (\vec q) = \sum_{n,n'}  M_{nn'}^{++} \Lambda_{nn'}^{++}(\vec q) \sim M_{n_1n_1}^{++} \Lambda_{n_1n_1}^{++}(\vec q)
\\
&\Pi_- (\vec q) = \sum_{n,n'}  M_{nn'}^{--} \Lambda_{nn'}^{--}(\vec q) \sim M_{n_2n_2}^{--} \Lambda_{n_2n_2}^{--}(\vec q) \label{eq:Pi-}
\end{align} 
Here, for simplicity, 
we kept only the predominant contribution which is from the $n_1$-th level in majority spin and $n_2$-th level in minority spin. And, we have defined $M_{nn'}^{ss'} = -\frac{f_{n}^{s}- f_{n'}^{s'}}{\epsilon_{n}^{s} - \epsilon_{n'}^{s'}}$, and the structure factor
\be\label{eq:def_Lambda}
\Lambda_{nn'}(\vec q) = \sum_{mm'}\int d\vec r\psi^*_{nm}(0)\psi_{nm}(\vec r)\psi^*_{n'm'}(\vec r)\psi_{n'm'}(0) e^{i\vec q\cdot \vec r}.
\ee  
What is the momentum of the leading density wave instability?
This can be answered by solving the maximum of the structure factors $\Lambda_{nn}(q)$. Plugging in the LL  wavefunction, using results in Ref.\cite{rajagopal1991linearized} and following the exactly same procedure as we did for SC in main text (from Eq.\eqref{eq:Delta1} to Eq.\eqref{eq:def_I}), we find the structure factors are given by
\begin{align}
\Lambda_{nn}(q) &=\lp\frac{1}{2\pi \ell_b^2}\rp^2 \int d^2\vec r \lb L_{n}\lp \frac{r^2}{2\ell_b^2} \rp \rb^2 e^{-r^2/2\ell_b^2} e^{i \vec q \cdot \vec r } \nonumber \\
&=  \frac{1}{2\pi \ell_b^2}  \frac{1}{\pi}\int r' d r' d\theta  \lb L_{n}\lp r'^2 \rp \rb^2 e^{-r'^2} e^{i \sqrt{2} \ell_b q r' \cos\theta } \nonumber\\
& = \frac{1}{\pi \ell_b^2} I_{nn}(x), \quad x = \sqrt{2} \ell_b q.\label{eq:Lambda}
\end{align}
The function $I_{nn}(x)$ is same as defined in main text Eq.\eqref{eq:def_I}. We calculate $I_{nn}(x)$ numerically and show it in Fig.\ref{fig:I_nn}. As shown in the figure, this function has a maximum of $I_{nn}(x=0) = 1/2$. Therefore, we conclude that the density wave instability is strongest at $q=0$.

However, the $q=0$ channel is essentially a uniform detunning field $d$ that originates from interaction. 
This observation implies that, before carrying out the analysis on all kinds of instabilities (including SC and density waves), we should first solve $d$ correctly by accounting for the impact of interaction $V$. In fact, as we will see below, finite-momentum density wave will no longer occur when detuning $d$ is accounted for. 

Motivated by this, we organize the remainder of our analysis as follows: In the Appendix.~\ref{app:detuning}, we will evaluate the detuning field $d$, and then plug it back into to Eq.\eqref{eq:Tc_d>>Tc} and Eq.\eqref{eq:intraflavor_dw_result1} to see if SC can occur. There, we will see that SC is usually killed by detunning, so long as pairing interaction $g$ is weaker than repulsion $V$. 
In Appendix.~\ref{app:density wave}, we show that density wave instabilities cannot occur in this model when detuning $d$ is correctly accounted for. 

\section{Detuning $d$ and the requirement for SC to occur in flat LL}\label{app:detuning}
In this appendix, we calculate the value of $d$. We will see, for temperature below an energy scale set by $V$: $T_V = \frac{V}{4\pi \ell_b^2}$, $d$ has a lower bound which is also $O(T_V)$.
Such $d$ turns out to be too large for SC to survive.

\begin{figure}
    \centering
    \includegraphics[width=0.99\linewidth]{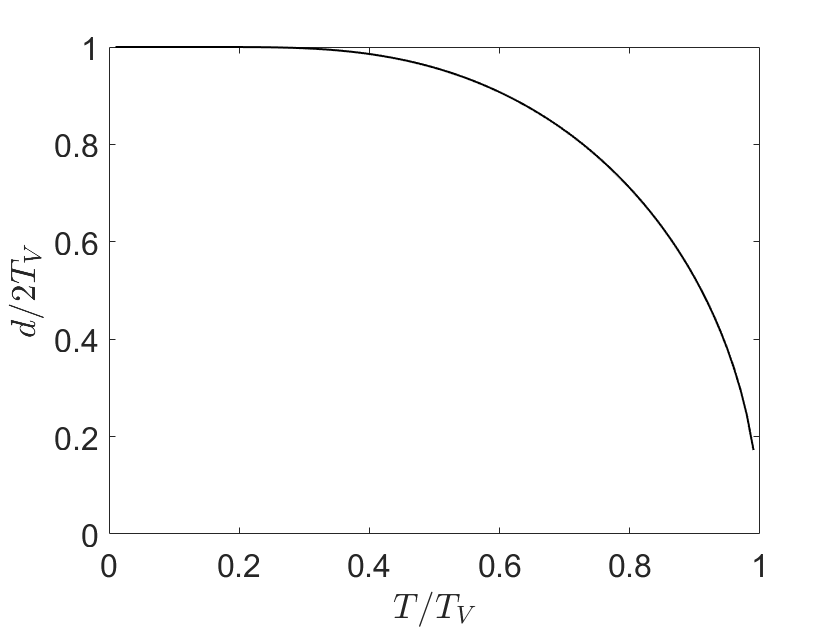}
    \caption{Self-consistent solution of detuning $d$ as a function of temperature $T$, obtained by numerically solving Eq.\eqref{eq:self-consistency_for_d_numeric}. Here we only present the result for the case of $c_1 = 0$.}
    \label{fig:d_vs_T}
\end{figure}

To start, we first analyze how detuning $d$ self-consistently adjust its value. The detuning $d$ is governed by the following self-consistency equation
\be\label{eq:d}
d =(\tilde{J}\mod\omega_c), \quad\tilde{J} = J + V(\rho_+ -\rho_-)
\ee
To solve this self-consistency equation, we first focus on the regime of $d\gg T$. In this regime, as discussed above, the alignment of LLs is as sketched in Fig.\ref{fig1} a). Therefore, the density top LLs in two flavors are given by
\be\label{eq:n+-}
\rho_- = \lp n_2-1+\frac{1}{e^{d/T}+1}\rp \frac{1}{2\pi \ell_b^2}, \quad\rho_+ = \rho-\rho_-,
\ee
where $\rho$ represents the total carrier density in the two levels, which is controlled extrinsically gate voltage. The quantity $\frac{1}{2\pi \ell_b^2}$ represents the carrier density in a fully filled LL. Plugging Eq.\eqref{eq:n+-} into Eq.\eqref{eq:d} and assuming $\omega_c\gg \frac{V}{2\pi \ell_b^2}$, we find
\be
 d = c_0 + \frac{V}{2\pi \ell_b^2} \tanh\lp \frac{d}{2T}\rp 
\ee
Here, $c_0$ absorbs all terms that are controlled externally $c_0 = ((J+\frac{V}{2\pi \ell_b^2}(n_1-n_2) + V\rho)\mod \omega_c)$. Equivalently we can rewrite it as
\be\label{eq:self-consistency_for_d_numeric}
 \frac{d}{2T_V} + c_1 = \tanh\lp \frac{d}{2T}\rp 
\ee
with $c_1= -c_0/2T_V$, $T_V = \frac{V}{4\pi \ell_b ^2}$. Below we analyze the solution of this self-consistency equation:
\begin{enumerate}
    \item for $T>T_V$, at a given $c_1$, self-consistency equation has at most one solution. 
 At $c_1=0$, the solution is $d=0$.
    Therefore, in this regime, $d$ can vanish upon tunning the value of $J$.
    \item For $T<T_V$, $|d|$ is no longer allowed to vanish --- it is always finite, and is minimized when $c_1=0$. Focusing on the case of $c_1=0$, we show $d$ as a function of $T$ in Fig.\ref{fig:d_vs_T}. We see that, $d$ scales as $\sqrt{T_V-T}$ right below $T=T_V$, and saturates at $d_{\rm min} = 2T_V$ at $T=0$. 
\end{enumerate}

Using the value of $d$ obtained above, we proceed to analyze the condition of for SC to occur. Recalling from our analysis below Eq.\eqref{eq:Tc_d>>Tc}, to allow SC, both detunning $d$ and temperature $T$ need to be sufficiently low, at least lower than $T_c^{(0)}$. However, it is not always possible to simultaneously tune these two quantities to this regime. As shown in Fig.\ref{fig:d_vs_T}, one cannot simultaneously make $d$ to $T$ much smaller than $T_V$. When $T\ll T_V$, $d\sim O(T_V)$, and vice versa. Therefore, SC can survive only if $T_c^{(0)}> T_V$, so that even though $d$ and $T$ are comparable to $T_V$, both of them are simultaneously below $T_c^{(0)}$.
This regime can only be achieved in system with a strong pairing interaction, such that $0.17g\gtrsim V$, but this is rare in natural systems. One possible way to achieve this condition is to screen the repulsive interaction $V$ by gates.

\section{Does density wave instabilities occur in flat LL?}\label{app:density wave}
In this appendix, we derive the condition of intra-flavor density wave instability, and show that, the presence of detuning $d$ will always kill the density waves instabilities in flat LL models.

To derive the onset condition for intra-flavor density wave instabilities, we consider a small change in carrier densities of two flavors $\delta \rho_\pm (r) = \sum_{q\neq 0} \delta \rho_\pm(q) e^{iqr}$. This change of density should shift the energies in two flavors. These energy shifts are given by
\begin{align}
\delta u_{+}(q) = V \delta \rho_-(q) + \kappa (\delta \rho_+(q) +\delta\rho_-(q)) 
  \nonumber\\
\delta u_{-}(q) = V \delta \rho_+(q) + \kappa (\delta \rho_+(q) +\delta\rho_-(q))
\label{eq:intraflavor_dw_1}
\end{align}
where $\delta u_\pm$ is the change of energy in $\pm$ flavor (relative to Fermi level) due to the charge density $\delta \rho_\pm$. 
The positive-valued quantity $V$ represents a contact repulsion. The first term is due to the direct repulsion between two flavors. For contact repulsion, the direct repulsion inside each flavor is canceled by exchange. The second term represents the pinning potential for the total charge, which is achieved by gate. In realistic system, $\kappa$ is the inverse of capacitance, which we take to be very large $\kappa \gg V$.  Meanwhile, $\delta \rho_\pm$ is associated with $\delta u_\pm$ through 
\be\label{eq:intraflavor_dw_2}
\delta \rho_\pm(q) = -\Pi_\pm (q)\delta u_\pm(q) 
\ee
where $\Pi_\pm (q)$ represents the total charge susceptibility in LLs, which we have calculated in Eq.\eqref{eq:Pi+} to Eq.\eqref{eq:def_Lambda}. 


Next, we proceed to solve Eq.\eqref{eq:intraflavor_dw_1}. Adding and subtracting the two lines of Eq.\eqref{eq:intraflavor_dw_1}, and
plugging in Eq.\eqref{eq:intraflavor_dw_2} yields
\begin{align}
(2\kappa+V)(\delta \rho_{+} + \delta \rho_{-})= -\frac{\delta\rho_+}{\Pi_+} - \frac{\delta\rho_-}{\Pi_-} 
\label{eq:intraflavor_dw_3}\\
(-\frac{1}{\Pi_+}+V) \delta\rho_{+} - (-\frac{1}{\Pi_-}+V) \delta\rho_{-}=  0 \label{eq:intraflavor_dw_4}
\end{align}
Due to large $\kappa$, Eq.\eqref{eq:intraflavor_dw_3} merely fix the total charge $\delta\rho_+ + \delta\rho_- = 0$. Plugging this into Eq.\eqref{eq:intraflavor_dw_4} yields the following condition
\be
V\tilde{\Pi}(q) = 1,\quad \tilde{\Pi}(q) = \frac{2\Pi_+(q)\Pi_-(q)}{\Pi_+(q) + \Pi_-(q)}.\label{eq:intraflavor_dw_result1}
\ee
As a reminder, around Eq.\eqref{eq:def_Lambda} we have shown that both $\Pi_+$ and $\Pi_-$ are maximized at $q=0$, as the structure factor is maximized at $q=0$. Therefore, below we first focus on $\Pi_+(0)$ and $\Pi_-(0)$. Based on our analysis from Eq.\eqref{eq:Pi+} to the text following Eq.\eqref{eq:def_Lambda}, $\Pi_+(0)$ and $\Pi_-(0)$ are given by
\begin{align}\label{eq:Pi}
\Pi_+ (0) = \frac{1}{2\pi \ell_b^2} M_{n_1n_1}^{++},\\
\Pi_- (0) = \frac{1}{2\pi \ell_b^2} M_{n_2n_2}^{--}, 
\end{align}
where, as a reminder, the quantity $M_{nn'}^{ss'}$ is defined as $M_{nn'}^{ss'}= -\frac{f_{n}^{s}- f_{n'}^{s'}}{\epsilon_{n}^{s} - \epsilon_{n'}^{s'}}$. For the case of $n'=n, s'=s$, $M_{nn'}^{ss'}$ can be calculated by taking derivative:
\be 
M_{nn}^{ss} = -\partial_{\epsilon_n} \lp \frac{1}{\exp(\epsilon_n/T) +1} \rp = \frac{1}{ 4T \cosh^2\lp\epsilon_n/2T \rp }.
\ee
In the presence of a detuning $d$, the Landau level in one of two flavors is away from Fermi level by $d$. Without losing generality, we assume the LL in $+$ flavor stay near Fermi level within a range of order $O(T)$, whereas the one in $-$ flavor is detunned from Fermi level by $d$. In this case, straightforward calculation gives
\be
\Pi_+(0) = \frac{1}{8 \pi \ell_b^2 T}, \quad  \Pi_-(0) = \frac{1}{8\pi \ell_b^2 T \cosh^2 (\frac{d}{2T}) },
\ee
Consequently, the spin-polarization susceptibility is 
\be
\tilde{\Pi}(0) = \frac{1}{8 \pi \ell_b^2 T } \frac{2 }{1+ \cosh^2 (\frac{d}{2T}) }
\ee
To see whether the density wave instability can occur, we calculate $V\tilde{\Pi}(0)$:
\be\label{eq:VPi}
V\tilde{\Pi}(0) = \frac{T_V}{ T } \frac{1}{1+\cosh^2(\frac{d}{2T})}
\ee
As a reminder, we have defined $T_V = \frac{V}{4\pi \ell_b^2 }$. Recall the relation between $d$ and $T$, $\frac{d}{2T_V} = \tanh(\frac{d}{2T})$, which is obtained by taking $c_1=0$ in Eq.\eqref{eq:d} in main text. Plugging this relation into Eq.\eqref{eq:VPi}, we find
\be
V\tilde{\Pi}(0) = \frac{x}{\lp 1+\cosh^2 x\rp \tanh{x} }, \quad x = \frac{d}{2T}
\ee
We find, the right-hand side is maximized at $x=0$. The maximal value of $V\tilde{\Pi}(0)$ is $1/2$. Therefore, we conclude, in  the flat-Landau level model, the density-wave instability will not be triggered in the presence of the self-consistent detuning. The basic picture is that the spontaneous generation of detuning $d$ occurs before any finite $q$ density wave instability. This pushes the LL away from the Fermi level, suppressing any further instability.

\section{Pairing in broadened Landau level}\label{app:Tc in broadened Landau level}
In this appendix, we switch to a modified model where LLs are broadened into minibands of width $W$ due to spatial inhomogenuity or disorder, as illustrated in Fig.\ref{fig1:combo}. 
For simplicity, we assume the correlation length of disorder potential is much greater than the magnetic length so that the wavefunction in the LL is not affected by disorder. In this setting, the analysis in previous sections that uses wavefunction in a uniform $b$ field as the electron wavefunction is still valid.



To generalize the analysis of the LL  problem (Eq.\eqref{eq:linearized_gap_eq_PDW}) to such finite-width miniband problem, all we need to do is to replace the summation over all degenerate states inside each LL by an integral 
a finite-width miniband.  Namely, in
Eq.\eqref{eq:linearized_gap_eq_PDW} we substitute 
\be
\frac{1}{(2\pi \ell_b^2)^2} \rightarrow \int d\epsilon^+d\epsilon^-  \nu_+ \nu_-\sim \nu_*^2 \int d\bar{\epsilon} d \epsilon' 
\ee
where we have defined $\bar{\epsilon} = \frac{1}{2}(\epsilon_{n_1}^+ + \epsilon_{n_2}^-)$, and $\epsilon' = \epsilon_{n_1}^+ - \epsilon_{n_2}^-$. Here, $\nu_\pm$ are the DOS in the relevant minibands in two spin flavors. 
For an estimate, we approximate  $\nu_\pm$ 
and $\nu_*$ by $\frac{1}{2\pi \ell_b^2 W}$, assuming that the Fermi level lies inside both $\pm$ minibands. Plugging this into the right-hand side (RHS) of Eq.\eqref{eq:linearized_gap_eq_PDW}, we find the integral over $\epsilon'$ simply yields a factor of $2W$ assuming $T_c \ll W$. Then, using $\nu_* W = \frac{1}{2\pi \ell_b^2} $, we estimate the RHS of Eq.\eqref{eq:linearized_gap_eq_PDW} as 
\be
{\rm RHS} \sim  2g\nu_* 
\int_{T_c}^{W} \frac{d\bar{\epsilon}}{\bar{\epsilon} } I_{n_1n_2}(p).
\label{eq:linearized_gap_eq_PDW_2}
\ee
Here, we have assumed that the cutoff scale of the pairing interaction $g$ is larger than $W$ so that the integral is cutoff at $W$. The value of the PDW momentum $P$ is the same as before. 
To maximize $T_c$, we choose $n_1=1$ and $n_2=0$, which gives the maximum value of $I(p)$ equal $\sim 0.17$. 

This yields 
\be\label{eq:enhanced Tc}
T_c = W e^{-1/0.34\lambda_*} 
,\quad
\lambda_* = g\nu_*
.
\ee
%
This result suggests that increasing 
$\omega_c$ , which enhances the density of states (DOS), can raise 
$T_c$ up to the point where 
$\lambda_*$ approaches order one. At this stage, the system transitions into the strong coupling regime, and 
$T_c$ saturates at $O(W)$. 
However, as we will see below,  the detuning parameter $d$ can set the upper bound of $T_c$. Namely, as usually Coulomb repulsion is stronger than pairing interaction, the dimensionless Coulomb coupling $V\nu_*$ reaches order-1 before $\lambda_*$. This leads to Stoner instability, which spontaneously generates a large detuning field $d$ even when external detuning is zero. As a result, one of two spin flavor will be gapped, suppressing SC. 


\section{Self-consistent treatment of the detuning energy $d$  in broadened LLs}\label{app:detuning_broad}
In this appendix we consider the self-consistent treatment of the detuning energy $d$ in the case when the LL has been broadened to acquire a width $W$. We write down the self-consistency equation governing detuning. First, the detuning energy $d$ is given by:
\be\label{eq:d}
d = J + V (
\rho_+- 
\rho_-) - \omega_c\Delta n,
\ee
where $\Delta n = n_1-n_2$ is the difference of band indices in two spins, $\rho_+$ represents the total carrier density in spin-majority flavor, whereas $\rho_-$ represents that of spin-minority flavor. 
The first term is an external spin-splitting due to exchange coupling to textured layer. The second term  describes the contribution of exchange energy to the spin splitting. The last term is an energy shift that translates spin splitting (which is the energy difference between the same minibands in two spin flavors) to detuning (which is the energy difference between $n_1$-th miniband of majority spin and $n_2$-th miniband in minority spin). On the other hand, the density difference is determined through
\be\label{eq:rho+-rho-}
\rho_+-\rho_- = \frac{\Delta n}{2\pi\ell_b^2}  + \nu_* d 
\ee
where the first term on right-hand side represents the density difference due to the difference in numbers of filled bands between two flavors, whereas the second term represents the density difference in conduction minibands. Here, we have assumed that $d$ is smaller than bandwidth $W$. Using these Eq.\eqref{eq:d} and Eq.\eqref{eq:rho+-rho-}, we find the following self-consistent solution of detuning
\be
d = \frac{d_0}{1-V \nu_*}
\ee
where we defined $d_0 = J +  \frac{V\Delta n}{2\pi \ell_b^2} -\omega_c \Delta n$, which represents the ``bare" detuning for the conducting minibands (top bands in  Fig.\ref{fig1:combo} a) 
of main text): it is the detuning calculated when ignoring the exchange energy of carriers in the conducting minibands. 

\section{Does density wave instabilities occur in broadened LL?}\label{app:density wave broad band}
In this appendix, we show that intra-flavor density wave instability does not occur in broadened LLs.

We still focus on the same situation as in Appendix.~\ref{app:detuning_broad} where Fermi levels lies inside the $n_1$-th miniband in majority spin and the $n_2$-th miniband in minority spin. The polarization function in two spin flavors $\Pi_s(q)$ ($s=\pm$) is given by
\begin{widetext}
\begin{align}\label{eq:Pi broad band1}
\Pi_s (\vec q) &\sim \sum_{mm'} -\frac{f(\epsilon_{nm})- f(\epsilon_{nm'})}{\epsilon_{nm} - \epsilon_{n'm'}} \int d\vec r\Psi^*_{nm}(0)\Psi_{nm}(\vec r)\Psi^*_{n'm'}(\vec r)\Psi_{n'm'}(0) e^{i\vec q\cdot \vec r}
\end{align}
\end{widetext}
where $n = n_1$ for $s=+$, $n = n_2$ for $s=-$, $f(\epsilon)=\frac{1}{\exp(\epsilon/T)+1}$ is the Fermi-Dirac distribution function, $\epsilon_{nm}$ and $\Psi_{nm}$ are the energy and wavefunction of $m$-th orbital in $n$-th miniband. Due to the presence of disorder-originated band broadening, $\epsilon_{nm}$ uniformly distribute within a window of $\lp \epsilon_n -\frac{W}{2},  \epsilon_n +\frac{W}{2} \rp$, where $\epsilon_n$ is the energy (measured from Fermi level) at the center of $n$-th miniband. 

As we have assumed in our model that the correlation length of disorder potential is much smaller than $\ell_b$, the energies of orbitals with different indices $m$ are not correlated. Therefore, we can perform an extra average over orbital indices in the integral in Eq.\eqref{eq:Pi broad band1}. Namely, $\Pi_s(q)$ can be rewritten as follows
\begin{widetext}
\begin{align}\label{eq:Pi broad band2}
\Pi_s (\vec q) & \sim \sum_{mm'} -\frac{f(\epsilon_{nm})- f(\epsilon_{nm'})}{\epsilon_{nm} - \epsilon_{n'm'}} \int d\vec r \langle \Psi^*_{nl}(0)\Psi_{nl}(\vec r)\Psi^*_{nl'}(\vec r)\Psi_{nl'}(0) e^{i\vec q\cdot \vec r}\rangle_{ll'} 
\end{align}
\end{widetext}
where $\langle ... \rangle_{ll'} = (2\pi \ell_b^2)^2 \sum_{ll'} ... $ is the average over all orbitals in $n$-th miniband. With this, we can proceed to calculate the two terms separately:
\begin{widetext}
\begin{align}
&\sum_{mm'} -\frac{f(\epsilon_{nm})- f(\epsilon_{nm'})}{\epsilon_{nm} - \epsilon_{nm'}} = \nu_*\\
&(2\pi \ell_b^2) \sum_{mm'}\int d\vec r\Psi^*_{nm}(0)\Psi_{nm}(\vec r)\Psi^*_{n'm'}(\vec r)\Psi_{n'm'}(0) e^{i\vec q\cdot \vec r} = (2\pi \ell_b^2)\Lambda_{nn}(q) = 2I_{nn}(\sqrt{2}\ell_b q).
\label{eq:sturcture factor broad band}
\end{align}
\end{widetext}
where in Eq.\eqref{eq:sturcture factor broad band} we have used Eq.\eqref{eq:Lambda}. As a result, the density wave susceptibility in broadened Landau minibands is given by
\be\label{eq:Pi_broad_result}
\Pi_s (\vec q) = 2\nu_* I_{nn}(\sqrt{2}\ell_b q)
\ee
where $n = n_1$ for $s=+$, $n = n_2$ for $s=-$. As a reminder, $I_{nn}(x)$ is maximized at $x=0$, $I_{nn}(0)=1/2$. Therefore, the maximum of $\Pi(q)$ is $\Pi(0) = \nu_*$. Plugging into Eq.\eqref{eq:intraflavor_dw_result1} we find the maximum of density wave susceptibility is also $\tilde{\Pi}(0) =\nu_*$. This is in agreement with what we found in Sec.\ref{app:detuning_broad}, where a detuning instability was found to occur at a Stoner threshold $V\nu_*=1$. We note that the result in Eq.\eqref{eq:Pi_broad_result} holds for arbitrary values of $d$ so long as the Fermi level lies inside the miniband. 

The main conclusion  
of this analysis is that  
polarization function values $\Pi_s(q)$ at finite $q$ are always weaker than $\Pi(0)=\nu_*$. 
As a result, the finite-momentum density wave will never occur before reaching the Stoner threshold for $q=0$. If $V$ exceeds the Stoner threshold,  a detuning will be generated that will push the miniband away from Fermi level until the Fermi level lies outside the LL width $W$.  In this case, finite $q$ density wave instability would clearly not occur since the DOS is greatly reduced by detuning. The only instability that survives is SC pairing, because pairing a consequence of a logarithmic divergence rather than a Stoner type instability.

\end{document}